\documentclass[twocolumn,,amsmath,amssymb]{revtex4}

\usepackage{graphicx}
\usepackage{dcolumn}
\usepackage{bm}
\newcommand{\etal}{\emph{et al.}}
\newcommand{\be}{\begin{equation}}
\newcommand{\ee}{\end{equation}}
\newcommand{\bfig}{\begin{figure}}
\newcommand{\efig}{\end{figure}}

\begin{document}      

\title{Observation of an edge supercurrent in the Weyl superconductor MoTe$_2$.
} 
\noindent

\author{Wudi Wang$^1$}
\author{Stephan Kim$^1$}
\author{Minhao Liu$^{1,**}$}
\author{F. A. Cevallos$^2$}
\author{R. J. Cava$^2$}
\author{N. P. Ong$^{1,*}$}
\affiliation{
Department of Physics$^1$ and Department of Chemistry$^2$, Princeton University, Princeton, NJ 08544
}

\date{\today}      
\begin{abstract} 
Edge supercurrents in superconductors have long been an elusive target. Interest in them has reappeared in the context of topological superconductivity. We report the observation of a robust edge supercurrent in the Weyl superconductor MoTe$_2$. In a magnetic field $B$, fluxoid quantization generates a periodic modulation of the edge condensate observable as a ``fast-mode'' oscillation of the critical current $I_c$ versus $B$. Remarkably, the fast-mode frequency is distinct from the conventional Fraunhofer oscillation displayed by the bulk supercurrent. We confirm that the fast mode frequency increases with crystal area as expected for an edge supercurrent. In addition, weak excitation branches are resolved which display an unusual broken symmetry. 
\end{abstract}

\maketitle      
In topological superconductors, attention has focused on topological edge states that carry excitations which are unpaired ~\cite{FuKane,Zhang,FuBerg}. A fundamental question is whether an edge supercurrent, distinct from the bulk supercurrent, can also exist. We report evidence for an edge supercurrent in the Weyl semimetal MoTe$_2$. The premise is that, if the edge condensate is sufficiently decoupled from the bulk condensate, fluxoid quantization periodically modulates the edge superfluid kinetic energy as the magnetic field $B$ is varied. We observe the modulation as a fast oscillation of the critical current $I_c$ vs. $B$. Tests confirm that the oscillations arise from a robust edge supercurrent. We also observe a weak Fraunhofer diffraction pattern associated with the bulk supercurrent, which is decoupled from the edge supercurrent.

The pairing of Weyl fermions has attracted considerable theoretical interest~\cite{Uchida,Aji,Burkov,ZhouWang,Madsen,Haldane}. To date, however, the only known Weyl superconductor is $\gamma$-MoTe$_2$~\cite{Soluyanov,Wang}, with a critical temperature $T_c\sim$ 100 mK at ambient pressure~\cite{Felser}. 

We contacted exfoliated crystals of $\gamma$-MoTe$_2$ (thickness $d$ = 60-120 nm) using evaporated Au probes (Tabel I and Sec. S1 in \cite{SI}). With the temperature $T$ fixed at 20 mK, we measured the differential resistance $dV/dI$ vs the bias current $I$ at selected $B$. The set of $dV/dI$ traces (100-200) are then represented in a color map of $dV/dI(B,I)$ in the $B$-$I$ plane. (Our experiment is distinct from proximity experiments~\cite{SI} in which supercurrent is injected from superconducting Al into graphene~\cite{Pablo} or HgTe/CdTe quantum wells~\cite{Yacoby,Delft}). 

\begin{figure*}[t]
\includegraphics[width=14 cm]{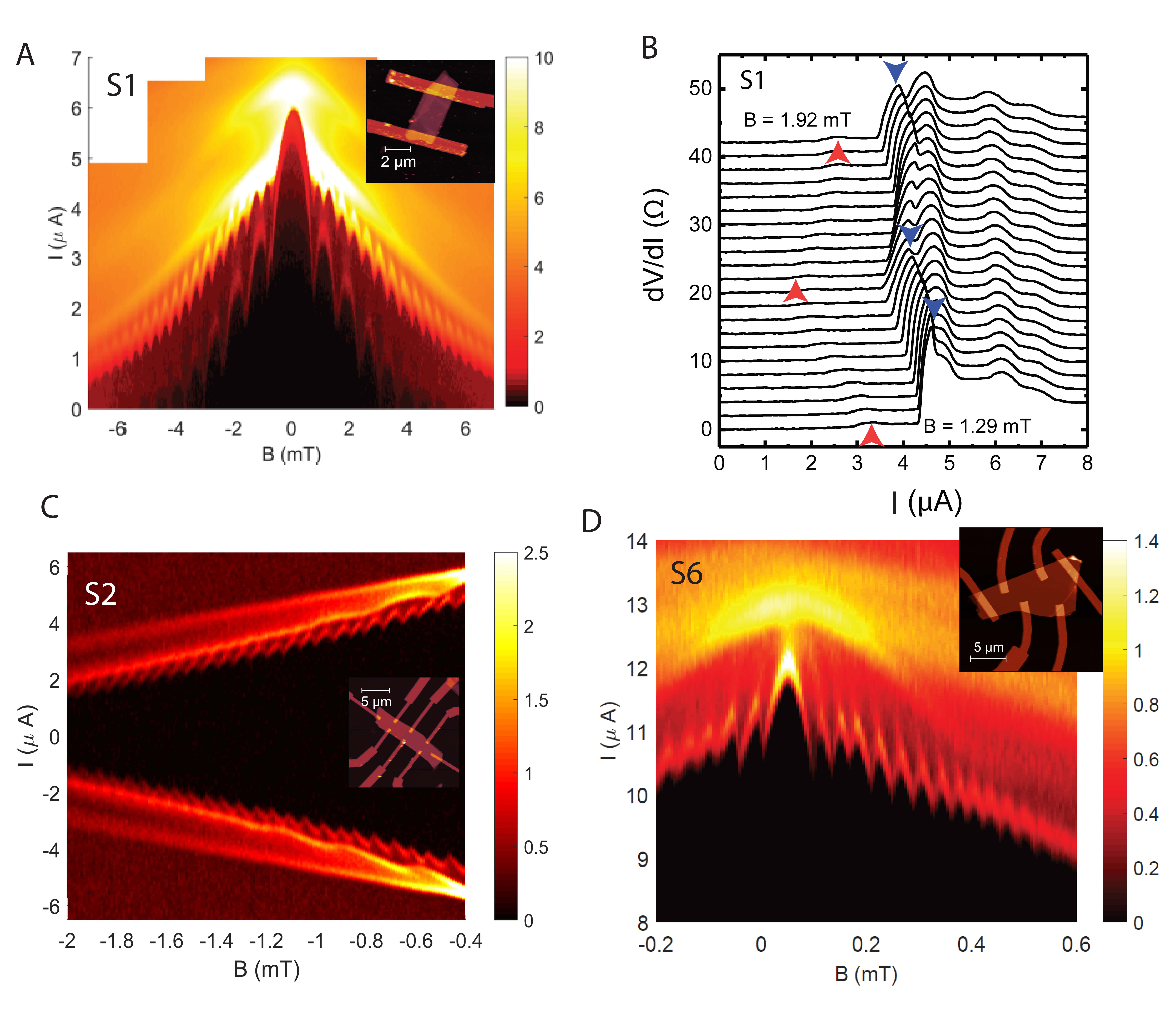}
\caption{\label{figS1S2} 
Color maps of the differential resistance $dV/dI$ vs. $I$ and $B$ in the Weyl superconductor MoTe$_2$ taken at 20 mK. In Sample S1 (Panel A), 2 oscillation modes are resolved. The fast mode, arising from amplitude modulation of an edge supercurrent, is observed as the scalloped boundary of the low-dissipation region (the scale bar shows $dV/dI$ in $\Omega$). The slow mode, associated with the bulk supercurrent, displays the usual Fraunhofer diffraction pattern. Panel B displays 22 traces of $dV/dI$ vs. $I$ (shifted for clarity) taken in S1 in steps of 30 $\mu$T starting at 1.29 mT. Prominent peaks (blue arrows) track the fast mode while the weak peaks (red arrows) track the slow mode. In large-area crystals (S2 and S6 in Panels B and C, respectively), the fast mode is strikingly evident, whereas the slow mode is unresolved. Insets show the Au contacts evaporated on each crystal. 
}
\end{figure*}

In a conventional superconductor, the exponential decay of flux precludes oscillatory behavior versus $I$ or $B$. By contrast, the color map in MoTe$_2$ (Fig. \ref{figS1S2}A, Sample S1) reveals a critical current $I_c(B)$ that oscillates with a scalloped profile which we call the fast mode. In addition, there exists a slow mode that arises from Fraunhofer diffraction. Panel B displays the traces of $dV/dI$ within a field interval comprising 2 periods of the fast mode. The large peaks (blue arrows) trace out the scalloped boundary, whereas the weaker peaks (red arrows) trace out the Fraunhofer diffraction pattern. In the color maps for two large-area samples S2 and S6 (Panels C and D, respectively), the fast mode is strikingly evident as the scalloped boundary surrounding the entire dissipationless region, whereas the slow mode is unresolved. We express the fast-mode frequency $f_1= 1/\Delta B_1$ (with $\Delta B_1$ the period) as a flux-penetration area $A_\phi \equiv f_1\phi_0$ ($\phi_0$ is the superconducting flux quantum).

The slow mode displaying the familiar Fraunhofer diffraction pattern reflects phase winding of the bulk supercurrent $J^b_s$, at a frequency $f_2 = 1/\Delta B_2$ that is not sensitive to the crystal area $A_{phys}$ (Fig. S3 in \cite{SI}). The conditions favoring observation of the slow mode (Fig. S5 in \cite{SI}) or the fast mode (Fig. S6) are described in Sec. S3 in \cite{SI}. 

\begin{figure*}[t]
\includegraphics[width=16 cm]{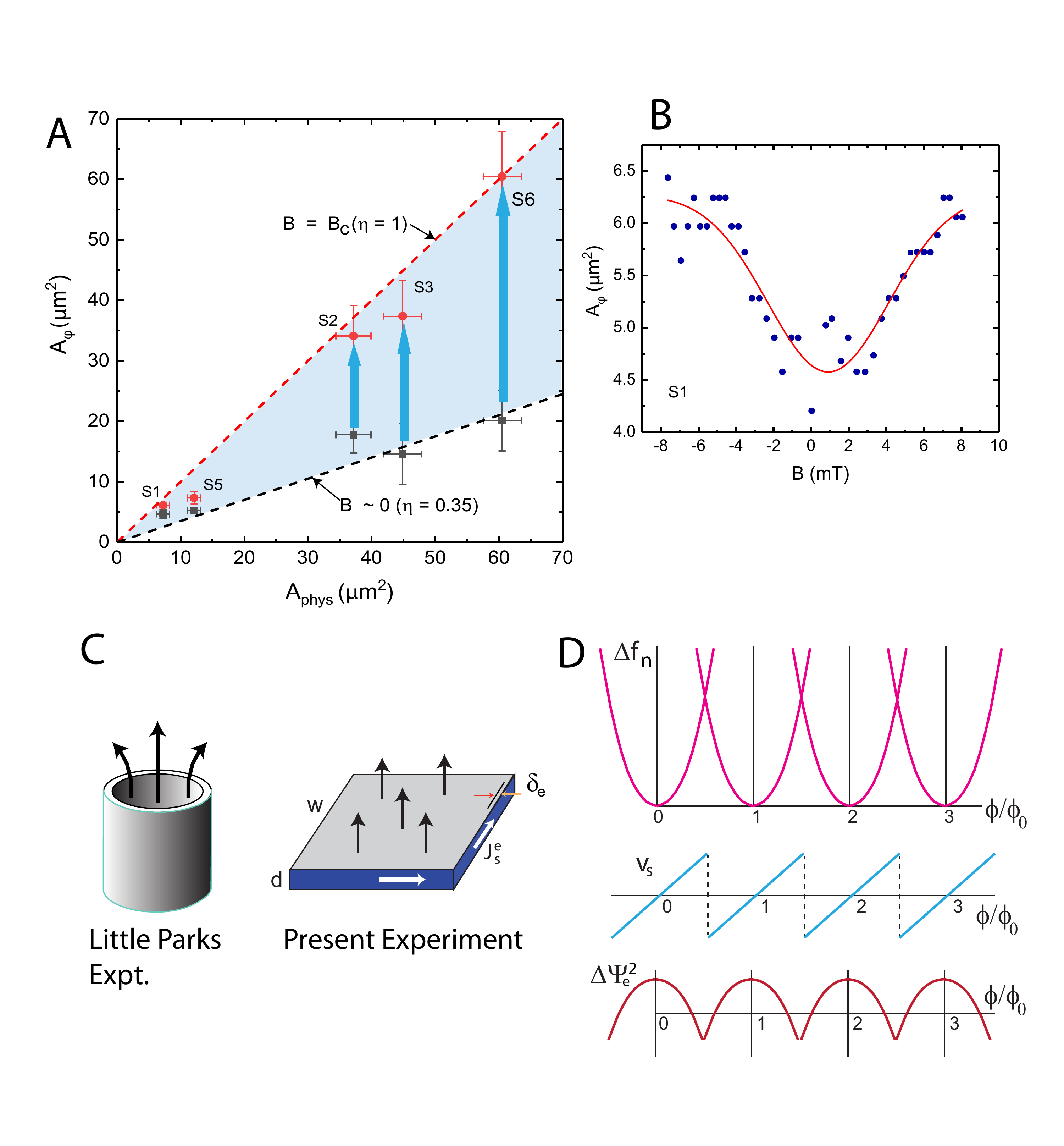}
\caption{\label{figArea} 
Area scaling, frequency chirp and scalloped profile. Panel A: Variation of the flux penetration area $A_\phi = \eta(B) A_{phys}$ in 5 samples, where $\eta(B)$ is the fraction of flux penetration in field $B$. In weak $B$, the data (black symbols) fall on the line with $\eta(B\sim 0) = 0.35$ (black dashed line). As $B\to B_c$, $\eta(B)$ in each sample increases towards 1 (broad arrows). In Panel B, the increase in $A_\phi$ vs. $B$ (in S1) saturates as $B\to B_c$ (red curve is a Gaussian fit). 
Panel C: Sketch of fluxoids (black arrows) trapped in a superconducting cylinder in the Little Parks experiment~\cite{Little} (left) and by the edge supercurrent $J^e_s$ (white arrows) in MoTe$_2$ (right). The width $\delta_e$ of $J^e_s$ is shown. Panel D: Changes in the superfluid kinetic energy lead to a set of branches of the free energy $\Delta f_n$, each centered at $\phi=n\phi_0$. Jumps between intersecting branches result in a sawtooth profile for ${\bf v}_s$ and oscillations in the edge condensate amplitude squared $\Delta\Psi_e^2$, observed as a characteristic scalloped boundary in the critical current $I_c(B)$.
}
\end{figure*}

Hereafter, we focus on $f_1$ to show that the fast mode originates from an edge supercurrent $J^e_s$.
Figure \ref{figArea}A shows that $f_1$, represented by $A_\phi$, scales as $A_\phi = \eta(B)A_{phys}$ across 5 samples. The fraction $\eta(B)$ expresses the degree of flux penetration. In the plot, the black symbols and black dashed line refer to the weak-field limit ($B<$1 mT). Already in this limit, $A_\phi = f_1\phi_0$ scales linearly with $A_{phys}$ with $\eta(B\sim 0) \simeq 0.35$.

Inspection of $f_1$ reveals that it increases gradually with $B$. This chirp effect reflects increasing flux penetration (on the scale of the Pearl length $\Lambda = 2\lambda^2/d$, where $\lambda$ is the London length). As indicated by the broad arrows and the red symbols, $A_\phi$ in each sample increases monotonically towards its physical area $A_{phys}$ as $B\rightarrow B_c$ (the critical field). The plot of $f_1$ vs. $B$ in Fig. \ref{figArea}B shows that it saturates as $B\rightarrow B_c$ so that $\eta(B\to B_c)\to 1$ but does not exceed 1. The partial screening implies that $J^b_s$ is not confined to a monolayer, but extends over the entire crystal volume.

Figure \ref{figArea}A shows that $f_1$ accurately tracks the flux quanta as $A_{phys}$ is increased 9-fold at fixed $B$, and also as $B\to B_c$ at fixed $A_{phys}$. Both trends suggest fluxoid quantization within a closed loop defined by $J^e_s$. We assume $J^e_s$ flows along the side wall (of depth $d$) encircling the crystal, with a width $\delta_e$ (see Fig. \ref{figArea}C), which we now estimate.
A finite $\delta_e$ leads to a spread in the area $\Delta A_\phi = \delta_e L_p$ and a phase uncertainty $\delta\varphi = 2\pi (\delta_e L_p/\phi_0)B$, where $L_p$ is the crystal perimeter. Complete dephasing of the fast mode occurs (at the dephasing field $B_d$) when $\delta\varphi\to\pi$. This yields $\delta_e = \phi_0/(2B_d L_p)$. From the observed $B_d=$ 9 mT in S1 (1.5 mT in S2), we find $\delta_e <$ 10 nm 
($\delta_e\sim$ 1/200 of the crystal width $w$). 

To make the case for fluxoid quantization (Sec. S5 in \cite{SI} for details), we assume that the edge condensate is described by a Ginzburg Landau (GL) wave function ($\hat{\Psi}_e$) distinct from that describing the bulk ($\hat{\Psi}_b$). The quantization of fluxoids within an enclosed area causes the edge superfluid velocity ${\bf v}_s$ to vary as $v_s = (2\pi\hbar/m^*L_p)(n-\phi/\phi_0)$ with $m^*$ the GL mass and $n\in {\cal Z}$~\cite{SI}. This leads to a set of free-energy branches $\Delta f_n(\phi)$ each centered at $\phi = n\phi_0$ (Fig. \ref{figArea}D). At an intersection, the system jumps between branches, leading to a sawtooth profile for $v_s(\phi)$. The result is a characteristic scalloped profile for the square of the wave function amplitude $\Psi_e^2\equiv |\hat{\Psi}_e|^2$ which we write as (Sec. S5 in \cite{SI})
\be
\frac{\Delta \Psi_e^2}{\Psi_e^2} = -{\cal P}\left(n-\frac{\phi}{\phi_0}\right)^2, 
\quad (n-\frac12 < \frac{\phi}{\phi_0} < n+\frac12),
\label{DPsi}
\ee
with the prefactor ${\cal P} = (2\pi\xi)^2/L_p^2$ where $\xi$ is the GL length.  

In the classic Little Parks (LP) experiment~\cite{Little,Tinkham}, the relative change corresponding to Eq. \ref{DPsi} is observed as a shift $\delta T_c(\phi)$ very near $T_c$ (where $\Psi_b\to 0$). 
Our experiment, performed at $T\ll T_c$, falls in a different regime; to drive both $\Psi_e$ and $\Psi_b\to 0$, we apply $I$ close to $I_c$. The narrow width $\delta_e$ of the edge condensate $\hat{\Psi}_e$ renders it less susceptible than $\hat{\Psi}_b$ to field suppression as $I$ approaches the boundary $I_c(B)$. Hence the edge $J^e_s$ carries an increasing share of $I$. At the boundary, $I_c \sim \Psi_e^2$ acquires the profile in Eq. \ref{DPsi}, i.e. $\Delta I_c\sim\Delta \Psi_e^2$ (Eq. S14 in \cite{SI}).

Equation \ref{DPsi} predicts that the oscillation amplitude $\Delta I_c$ decreases steeply as $1/L_p^2$. We confirm that the observed decrease is consistent with the prediction (see Fig. S8 in \cite{SI}). The model also explains a striking observation. As seen in Samples S1, S2 and S6 in Fig. \ref{figS1S2}, the fast-mode minima occur high above the horizontal axis, $I=0$, whereas the slow mode minima in S1 (also V2 in Fig. S5 of \cite{SI}) reach nearly to zero. This occurs because the former arises from a weak modulation of the amplitude $\Delta \Psi_e^2$, whereas the latter derives from phase winding. 

\begin{figure*}[t]
\includegraphics[width=16 cm]{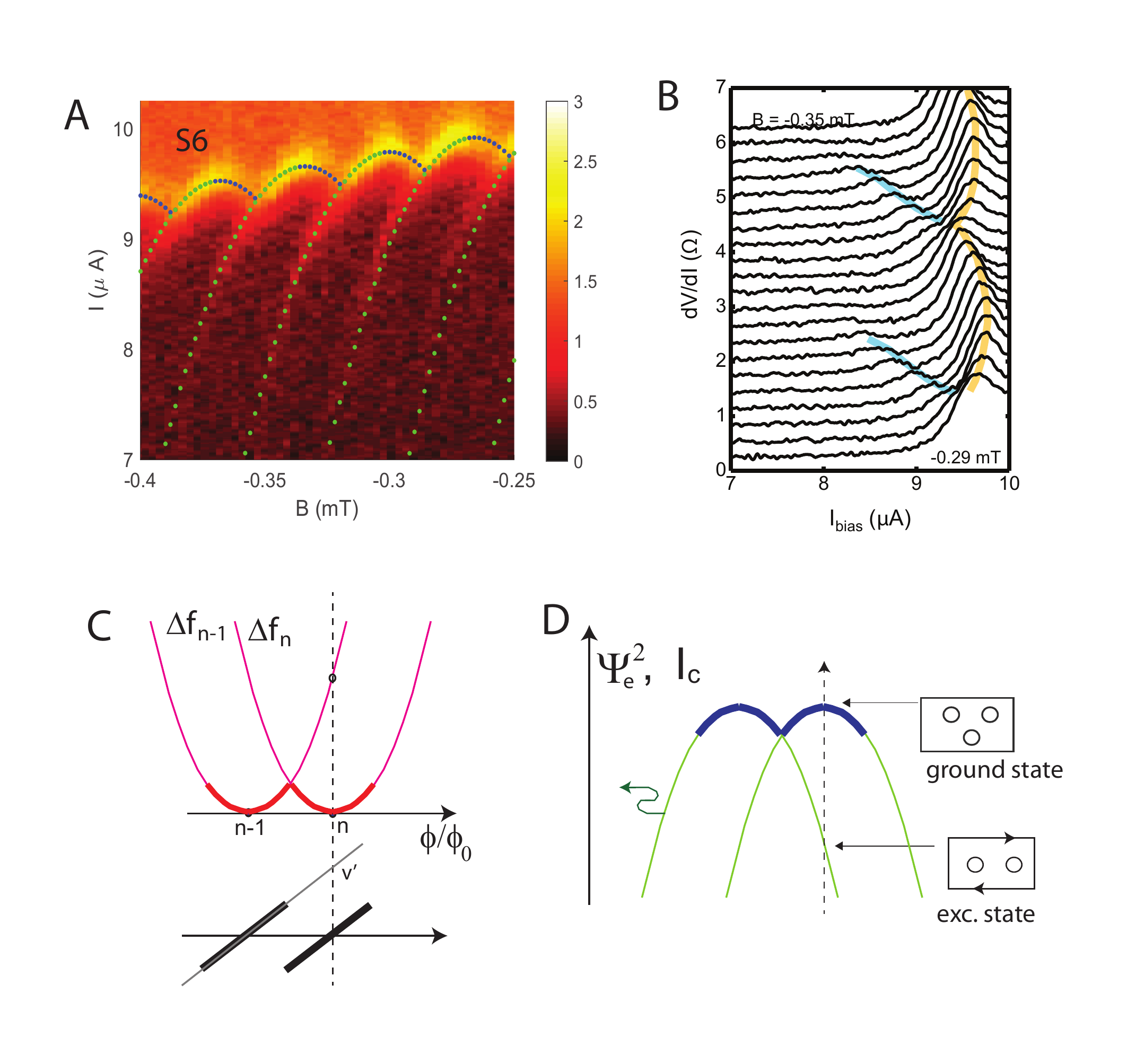}
\caption{\label{figExcitation} 
Emergence of excitation branches. Panel A: High resolution color map of $dV/dI$ curves showing weak excitation branches trailing from each minimum in the scalloped boundary. The data are obtained at 20 mK in S6 with $B<0$ and $I>0$. Green dots represent fits to Eq. \ref{DPsi}. Panel B displays 21 traces of $dV/dI$ vs. $I$ in the interval -0.29 $<B<$ -0.35 mT (shifted vertically for clarity). The scalloped boundary (yellow curve) is traced by the large peak. At each cusp, a weak peak emerges and branches off to the left to trace out an excitation branch (blue curve). Panel C: Schematic plots of $\Delta f_n$ and $\Delta f_{n-1}$ (magenta parabolas), and the sawtooth profile of ${\bf v}_s$. The corresponding curves of $\Psi_e^2$ are plotted in Panel D (green parabolas). Bold blue arcs represent the scalloped boundary of $I_c(B)$. With $\phi$ fixed at $n\phi_0$ (dashed lines), the system occupies the lowest energy branch with $n$ = 3 fluxoids and $v_s = 0$. When $I$ is scanned at fixed $B$, the excited state (with 2 fluxoids and a large $v_s$) is encountered at a current smaller than $I_c(B)$. This is observed as the excitation branch. 
}
\end{figure*}

Next, we turn to a feature not observed in the LP experiment. The set of $\Delta f_n$ curves suggest that, at low $T$, it is possible to detect excited states. Using high-resolution scans, we have resolved weak excitation branches trailing from the scalloped boundary (Fig. \ref{figExcitation}A). As shown by the green dots in Panel A, the branches fit well to Eq. \ref{DPsi}. The excitations are also directly visible in individual traces of $dV/dI$ vs. $I$ (Fig. \ref{figExcitation} B). The large peak traces out the arcs of the scalloped boundary (yellow curve). At the cusp, a small peak (20-30$\times$ weaker in strength) emerges and traces out an excitation branch (blue curve). These excitations are also seen in S2 (Fig. S9 in \cite{SI}).

Our scenario for the excitation branch is sketched in Figs. \ref{figExcitation}C and \ref{figExcitation}D. When $\phi$ is fixed at $n\phi_0$ (dashed line), the system lies at the minimum of $\Delta f_n$ (magenta curves). Accordingly, the ground state has ${\bf v}_s = 0$ with $n$ fluxoids. The intersection of the branch $\Delta f_{n-1}$ with the dashed line defines an excited state with $n-1$ fluxoids and a large superfluid velocity $v'$. Expressed in terms of $\Psi_e^2$ (equivalently $I_c$), the free energy minima become the scalloped boundary (bold curves in Fig. \ref{figExcitation}D). As $I$ is increased (along the dashed line), we encounter the excited state at a value of $I$ ($<I_c(B)$) that varies with $\phi$ as in Eq. \ref{DPsi}.

\begin{figure*}[t]
\includegraphics[width=18 cm]{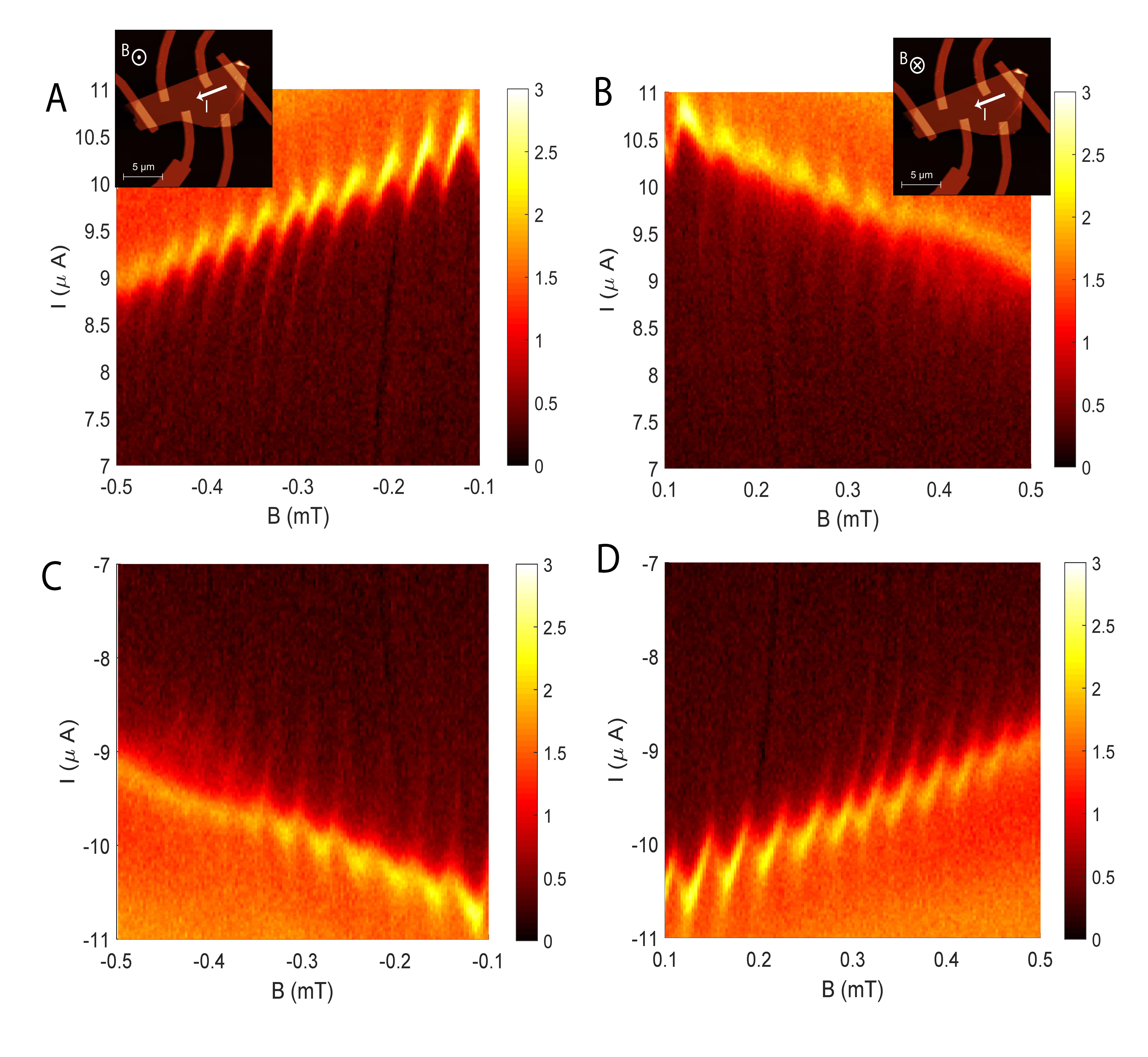}
\caption{\label{figS6} 
Symmetry breaking in the excitation branches in Sample S6. In Panel A the color map is measured at 20 mK with $B<0$ (out of page in the inset) and $I>0$ (flowing right to left). As described in Fig. \ref{figExcitation}A, the observed excitation branches flow to the left (increasing $|B|$). If $B$ is reversed keeping $I>0$ (see inset), the excitation branches flow left (Panel B), towards decreasing $|B|$, although the pattern is less sharply resolved than in A. If $B<0$ and $I<0$ (Panel C), the color map is similar to that in Panel B except for the reversal in $I$. Finally, for $B>0$ and $I<0$ (Panel D), we recover the color map in Panel A. The symmetry follows the sign of the product $I\cdot B$. For $I\cdot B>0$ (Panels B and C), the branches flow towards decreasing $|B|$ whereas for $I\cdot B<0$, the flow is towards increasing $|B|$ (A and D). In each panel, the patterns are non-hysteretic and independent of field-sweep direction. Sample S2 shows a similar symmetry breaking~\cite{SI}.
}
\end{figure*}

Lastly, we discuss an interesting asymmetry exhibited by these branches. In Fig. \ref{figExcitation}A, branches that flow outwards (towards increasing $|B|$) are observed while branches flowing inwards are conspicuously absent. As shown in Fig. \ref{figS6}, the flow direction is sensitive to the signs of $I$ and $B$. Panel A, with outflowing branches, is the situation already discussed ($I>0$, $B<0$). When we reverse the sign of $B$ (keeping $I>0$, Panel B), the branches flow towards decreasing $|B|$ although less clearly resolved. Likewise, in Panel C ($I<0$, $B<0$), the flow is towards decreasing $|B|$. Finally, with $I<0$ and $B>0$ (D), we recover the pattern in Panel A. The branches flow outwards if the product $I\cdot B<0$ (Panels A and D) whereas they flow inwards if $I\cdot B>0$ (B and C). The pattern favors one circulation of ${\bf v}_s$ over the other (but respects time-reversal invariance). These symmetry patterns, lying beyond the scenario discussed, require the role of spin-orbit coupling and other topological properties of the edge modes to be better understood. 

Aside from the symmetry breaking, the mechanism that protects the edge condensate against hybridization with the bulk, and the role played by hinge states~\cite{Bernevig} are issues under active investigation. More broadly, this method may be extended to explore other topological superconductors~\cite{FuKane,Zhang,FuBerg} and chiral superconductors~\cite{Ivanov,Moler}.


\newpage

\vspace{1cm}\noindent
$^\dagger$Corresponding author's email: npo@princeton.edu\\
$^*$Present address of M.L.: Zitan Technologies, Tahoe Blvd, Incline Village, NV 89450

\vspace{5mm}\noindent
{\bf Acknowledgments}\\
We thank B. A. Bernevig, B. I. Halperin, A. Yacoby and A. Yazdani for valuable discussions. The research was supported by the U.S. Army Research Office (W911NF-16-1-0116). The dilution refrigerator experiments were supported by the Department of Energy (DE-SC0017863). NPO and RJC acknowledge support from the Gordon and Betty Moore Foundation's Emergent Phenomena in Quantum Systems Initiative through Grants GBMF4539 (NPO) and GBMF-4412 (RJC). The growth and characterization of crystals were performed by FAC and RJC, with support from the National Science Foundation (NSF MRSEC grant DMR 1420541). 

\vspace{5mm}
\noindent
{\bf Supplementary Materials}\\
Supplementary Text\\
Figs. S1 to S9\\
Table I

\newpage


\end{document}